\documentclass[aps,prl,article,twocolumn,preprintnumbers,amsmath,amssymb,superscriptaddress]{revtex4-2}

\usepackage{graphicx}  
\usepackage{dcolumn}   
\usepackage{bm}        
\usepackage{amssymb}   
\usepackage{amsmath}
\usepackage{mathrsfs}
\usepackage{epigraph}
\usepackage{braket}
\usepackage{gensymb}
\usepackage{lmodern}
\usepackage{ tipa }
\usepackage{bbold}
\usepackage{esint}
\usepackage{mathdots}
\usepackage{xcolor}
\usepackage{appendix}
\usepackage{natbib}
\usepackage{multirow}
\usepackage{float}
\usepackage{physics}
\usepackage[normalem]{ulem}
\usepackage{times}

\usepackage{comment}
\usepackage[colorlinks=true, linkcolor=blue, urlcolor=blue, citecolor=blue]{hyperref}

\newcommand{\win} {{\omega_{\rm in}}}

\newcommand{\dw} {{\omega}}
\newcommand{\mpol} {{m\mkern-1mup}}

\begin{document}
\title{Witnessing Spin-Orbital Entanglement Using Resonant Inelastic X-Ray Scattering} 
\author{Zecheng Shen}\thanks{These authors contributed equally to this work.}
\affiliation{Department of Chemistry, Emory University, Atlanta, Georgia 30322, USA}
\author{Shuhan Ding}\thanks{These authors contributed equally to this work.}
\affiliation{Department of Chemistry, Emory University, Atlanta, Georgia 30322, USA}
\author{Zijun Zhao}
\affiliation{Department of Chemistry, Emory University, Atlanta, Georgia 30322, USA}
\author{Francesco A. Evangelista}
\affiliation{Department of Chemistry, Emory University, Atlanta, Georgia 30322, USA}
\author{Yao Wang}
\email[Correspondence should be addressed to \href{mailto:yao.wang@emory.edu}{yao.wang@emory.edu}]{}
\affiliation{Department of Chemistry, Emory University, Atlanta, Georgia 30322, USA}
\date{\today}
\begin{abstract}  
Entanglement plays a central role in quantum technologies, yet its characterization and control in materials remain challenging. Recent developments in spectrum-based entanglement witnesses have enabled new strategies for quantifying many-body entanglement in macroscopic materials. Here, we develop a protocol for detecting spin-orbital entanglement using experiment-accessible resonant inelastic x-ray scattering. Central to our approach is the construction of a Hermitian generator from measurable spectra, which allows us to compute the quantum Fisher information (QFI) available in spin-orbital systems. The resulting QFI provides upper bounds for $k$-producible states and thus serves as a robust witness of spin-orbital entanglement. To account for realistic experimental limitations, we further extend our framework to include relaxed QFI bounds applicable to measurements lacking full polarization resolution.
\end{abstract}

\maketitle

As research on quantum materials deepens, a wide range of many-body phenomena, which fundamentally depart from the single-particle band theory at the core of traditional solid-state physics\,\cite{keimer2017physics, basov2017towards}, have been discovered. These phenomena, ranging from high-$T_c$ superconductivity to quantum spin liquids, are driven not by individual electrons but by collective behavior across the entire system, governed by many-body entanglement. As a result, entanglement has become both a signature of quantum matter and a critical resource for emerging technologies in quantum computation, teleportation, and sensing\,\cite{guhne2005multipartite, brukner2006crucial, amico2008entanglement}. Quantitatively measuring and controlling entanglement in real materials has evolved into a new research frontier\,\cite{laurell2025witnessing}.

A major advance in recent years stems from the development of spectrum-based entanglement witnesses, which have enabled the detection of entanglement in macroscopic solid-state systems\,\cite{laurell2025witnessing, hauke2016measuring}. Neutron scattering experiments have successfully leveraged this approach to uncover multipartite spin entanglement in magnetic chains\,\cite{mathew2020experimental, laurell2021quantifying, menon2023multipartite}, quantum spin liquids\,\cite{scheie2024proximate}, and strange metals\,\cite{fang2025amplified, mazza2024quantum}. These measurements rely on a formal correspondence between the dynamical spin response and the quantum Fisher information (QFI), a metrological quantity that characterizes the magnitude of quantum fluctuations\,\cite{hyllus2012fisher,toth2014quantum}. By constructing QFI from spin operators and comparing the experimentally inferred values to those of Greenberger-Horne-Zeilinger states, one can perform rigorous Bell-type tests for entanglement.

Despite its broad utility, this approach is often built on the assumption of a two-level system, inherited from the qubit paradigm in quantum information\,\cite{costa2021entanglement}. However, this simplification may obscure the structure of entanglement in correlated materials, where spin, charge, and orbital degrees of freedom are intertwined\,\cite{davis2013concepts,oles2006spin,you2012neumann, gotfryd2020spin}. For instance, in transition-metal materials, the low-energy manifold frequently includes multiple nearly degenerate orbitals; in heavy-fermion compounds, strong spin-orbit coupling further complicates the picture. As a result, restricting spatial entanglement to the spin sector may underestimate or misrepresent the many-body entanglement. Furthermore, recent advances in x-ray scattering techniques now allow for simultaneous probing of spin, charge, and orbital excitations\,\cite{ament2011resonant, tohyama2015resonant,mitrano2024exploring,de2024resonant}, further motivating entanglement probes incorporating intertwined degrees of freedom.

In this Letter, we investigate spin-orbital entanglement in systems where spin and orbital degrees of freedom are intrinsically nonfactorizable. Because Hermitian orbital excitations are generally inaccessible in experiments\,\cite{ren2024witnessing}, we introduce a practical entanglement witness based on a pair of polarization-reversed resonant inelastic x-ray scattering (RIXS) measurements. This approach enables the construction of an effective Hermitian operator and its corresponding QFI metric to quantify joint spin-orbital excitations. We further examine both tight and relaxed bounds of this metric under different experimental conditions.

\begin{figure}
    \centering
    \includegraphics[width=\linewidth]{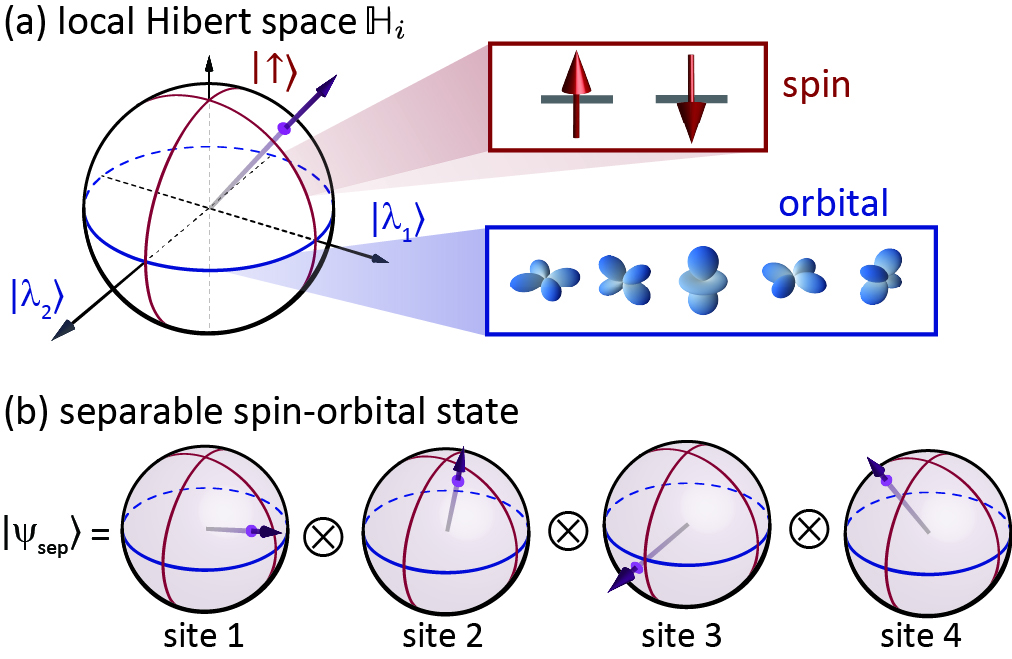}\vspace{-3mm}
    \caption{(a) Schematic illustration of the local Hilbert space $\mathbb{H}_i = \mathbb{H}_i^{\rm{spin}} \otimes \mathbb{H}_i^{\rm{orb}}$, with spin and orbital degrees of freedom at each site. (b) Example of a separable spin-orbital state.\vspace{-1mm}}
    \label{fig:cartoon}
\end{figure}

When focusing exclusively on spin excitations, the QFI is realized through the spin excitation operators, typically expressed in momentum space as $S_{\bm{q}}=\sum_{j} e^{i\bm{q}\cdot \bm{r}_j} S_{j}$. This formulation implicitly assumes that orbital degrees of freedom are either frozen or irrelevant for entanglement. To capture spin-orbital entanglement in quantum materials, we begin by formalizing the structure of the local Hilbert space. Unlike pure spin models, the local Hilbert space at each lattice site, constructed from local Wannier orbitals, is naturally extended as
\begin{align}
  \mathbb{H}_i
    =& \mathbb{H}_i^{\rm{spin}}\otimes \mathbb{H}_i^{\rm{orb}} \nonumber\\
    =& \operatorname{span}\{\ket{\uparrow},\ket{\downarrow}\}\otimes
       \operatorname{span}\bigl\{\ket{\lambda}\mid
         \lambda = 1,\dots,N_{\rm{orb}}\bigr\} .
\end{align}
As shown in Fig.~\ref{fig:cartoon}(a), this expanded Hilbert space incorporates both spin-$1/2$ and the full orbital manifold at site $i$, occupied by a single electron or hole. Here, $\ket{\lambda}$ denotes a local orbital basis, and $N_{\rm{orb}}$ is the number of active orbitals per site. While $N_{\rm{orb}}$ may, in principle, be infinite (spanning over core and unoccupied states), it is usually truncated to a physically relevant subspace composed of valence and conduction orbitals. For example, in transition-metal compounds such as cuprates or infinite-layer nickelates, the physically relevant space is the fivefold $3d$ manifold: $\{d_{x^{2}-y^{2}}, d_{3z^{2}-r^{2}}, d_{xy}, d_{xz}, d_{yz}\}$\,\cite{truncationNote}. 

A pure many-body state is said to be separable if it can be written as a direct product of local states, $\ket{\Psi_{\rm{sep}}} = \ket{\phi_1}\otimes\cdots\otimes\ket{\phi_N}$, where each $\ket{\phi_i}$ lies within the local Hilbert space $\mathbb{H}_i$ as illustrated in Fig.~\ref{fig:cartoon}(b). This formulation assumes that there is no inter-site tunneling and that local particle numbers are strictly conserved. If these assumptions are violated, a fermionic many-body formalism is required, where conventional mode-based witnesses become inadequate\,\cite{schliemann2001quantum, ghirardi2002entanglement, ghirardi2004general, amico2008entanglement, eckert2002quantum, kraus2009pairing, liu2025entanglement}. Within this framework, the notion of $k$-producibility extends naturally to the spin-orbital Hilbert space. A pure many-body state $\ket{\Psi_{k-\rm{prod}}}$ is $k$-producible if it factorizes into nonoverlapping partitions, each involving no more than $k$ sites:
\begin{equation}\label{eq:HilbertSpace}
\ket{\Psi_{k-\rm{prod}}} = \ket{\Phi_1}\otimes \ket{\Phi_2}\otimes\cdots\otimes \ket{\Phi_m} ,\quad|P_l| \le k .
\end{equation}
Here, each $\ket{\Phi_l}$ denotes an irreducible many-body state in partition $P_l$, with no more than $k$ spin-orbital degrees of freedom. The set of such states thus defines a bounded hierarchy that provide a natural basis for quantifying multipartite entanglement.

A central requirement for measuring QFI in solid-state systems is that it be constructed from the off-diagonal fluctuations of a Hermitian operator associated with the relevant mode\,\cite{hauke2016measuring, laurell2025witnessing,costa2021entanglement}. In practice, however, such Hermitian excitations are not always experimentally accessible, as elaborated below. This challenge is particularly pronounced when probing spin and orbital degrees of freedom,  for which RIXS has emerged as a powerful spectroscopic method. RIXS is especially effective due to its sensitivity to orbital excitations that are forbidden in conventional optical spectroscopy\,\cite{gel1994resonant,butorin1996low,ghiringhelli2004low,schlappa2018probing,li2021unraveling}. RIXS accesses these excitations indirectly via a two-step process: an incident x-ray photon excites the system into a short-lived intermediate state via a dipole excitation, and this state subsequently decays by emitting a second x-ray photon. The difference in energy and momentum between the incoming and outgoing photons encodes the dynamical response of valence electrons. The RIXS cross section follows the Kramers-Heisenberg formula\,\cite{ament2011resonant},
\begin{equation}\label{eq:RIXS}
    I({\bm{q}},\dw,\win) = \frac{1}{\pi}\rm{Im}\bra{\Psi_{\rm f}(\mathbf{q})}\frac{1}{\mathcal{H}-\textit{E}_G-\dw-\textit{i}0^+}\ket{\Psi_{\rm f}(\mathbf{q})} ,
\end{equation}
where the final-state wave function $\ket{\Psi_{\rm f}(\mathbf{q})}$ depends on the incident photon energy $\omega_{\rm{in}}$,
\begin{equation}\label{eq:finalState}
    \ket{\Psi_{\rm f}(\mathbf{q})} = \sum_{j}e^{i{\bm{q}} \cdot \bm{r}_j} \mathcal{D}_j^{(\epsilon_{\rm{s}})\dagger} \frac{1}{\mathcal{H}'-E_\text{G}-\omega_{\rm{in}}-i\Gamma}\mathcal{D}^{(\epsilon_{\rm{i}})}_j\ket{G} .
\end{equation}
Here, ${\bm{q}}$ and $\omega$ denote the momentum and energy transfers, respectively, and $E_G$ is the ground-state energy. The operator $\mathcal{D}^{(\epsilon)}_j =  \sum_{\alpha\beta} M_{\alpha\beta}(\epsilon) h^\dagger_{j\alpha}c_{j\beta}$ describes transitions between the core and valence levels, with $\epsilon=\epsilon_{\rm i/s}$ labeling the polarization of the incoming or outgoing photon. In this expression, $c_{j\beta}$ annihilates a valence hole, $h^\dagger_{j\alpha}$ creates a core hole at site $\bm r_j$, and $M_{\alpha\beta}(\epsilon)$ denotes the dipole matrix element between spin-orbital states indexed by $\alpha$ and $\beta$. (For the convenience of the latter examples, we employ the hole language in this Letter, but the framework also applies to electron langauge.) The intermediate Hamiltonian $\mathcal{H}'$ includes the presence of the core hole and its coupling to valence electrons, and the parameter $1/\Gamma$ sets the lifetime of the intermediate state.

Although the finite lifetime of the core hole distinguishes RIXS from conventional two-point dynamical correlators, a rigorous treatment of entanglement witnesses in the RIXS spectrum requires modeling the core-level Hamiltonian\,\cite{liu2025entanglement}. To focus on the interplay between spin-orbit entanglement and preserve the locality defined in Eq.~\eqref{eq:HilbertSpace}, we employ the ultrashort core-hole lifetime (UCL) approximation ($\Gamma \rightarrow \infty$), a widely used limit for capturing collective excitations\,\cite{ament2009theoretical,jia2016using} [see Supplemental Material(SM) for a discussion of finite lifetime effects \,\cite{SI_note}].
In this limit, the RIXS final-state wave function simplifies to $\ket{\Psi_{\rm f}(\mathbf{q})} \approx i\mathcal{T}_{{\bm{q}}}(\epsilon_{\rm i},\epsilon_{\rm s})\ket{G}/\Gamma$, where the scattering operator $\mathcal{T}_{{\bm{q}}}(\epsilon_{\rm i},\epsilon_{\rm s})$ captures polarization-resolved valence-band excitations:
\begin{align} \label{eq:scatteringOperator} 
    \mathcal{T}_{{\bm{q}}}(\epsilon_{\rm i},\epsilon_{\rm s}) &=  \sum_{j}e^{i{\bm{q}} \cdot \bm{r}_j} \mathcal{D}_j^{(\epsilon_{\rm{s}})\dagger}\mathcal{D}_j^{(\epsilon_{\rm{i}})} \nonumber\\
    &= \sum_{j,\alpha\beta} e^{i {\bm{q}} \cdot \bm{r}_j} T_{\alpha\beta}(\epsilon_{\rm i},\epsilon_{\rm s}) c_{j\alpha}^\dagger c_{j\beta}.
\end{align}
All spin-orbital excitations accessible by RIXS are embedded in the matrix $T_{\alpha\beta}(\epsilon_{\rm i},\epsilon_{\rm s}) = \sum_{\gamma}M^*_{\gamma\alpha}{(\epsilon_{\rm s})}M_{\gamma\beta}{(\epsilon_{\rm i})}$, which depends solely on the polarization configurations. 

In general, both $\mathcal{T}_{{\bm{q}}}$ and $T_{\alpha\beta}$ are non-Hermitian. Consequently, even in the UCL limit, the RIXS operator fails to satisfy the Hermiticity condition necessary for extracting QFI\,\cite{hauke2016measuring}, and this limitation becomes more severe when finite-lifetime corrections from Eq.~\eqref{eq:finalState} are included (see SM\,\cite{SI_note} and Ref.\,\cite{ren2024witnessing}). Accordingly, the zero-temperature spectral integral of RIXS,
\begin{equation}\label{eq:RIXSIntegral}
    \Gamma^2\!\int_{0}^\infty \mkern-4mu I({\bm{q}},\omega,\win) \mathrm{d}\omega = \langle \mathcal{T}^\dagger_{\bm{q}} \mathcal{T}_{\bm{q}} \rangle - \langle \mathcal{T}^\dagger_{\bm{q}} \rangle \langle \mathcal{T}_{\bm{q}} \rangle = \langle \mathcal{T}^\dagger_{\bm{q}} \mathcal{T}_{\bm{q}} \rangle_c ,
\end{equation}
\nocite{evangelista.2024.10.1063/5.0216512,ilias.2007.10.1063/1.2436882,liu.2007.10.1063/1.2710258,liu.2009.10.1063/1.3159445,saue.2011.10.1002/cphc.201100682,dyall.2007.10.1093/oso/9780195140866.001.0001,reiher.2014.10.1002/9783527667550}

cannot be interpreted as a genuine quantum fluctuation. Here, $\langle \cdots \rangle_c$ denotes the cumulant expectation value, and polarization dependence has been omitted for brevity. While it has been proposed that including anti-Stokes contributions might restore Hermiticity\,\cite{ren2024witnessing}, these processes are exponentially suppressed at low temperatures and typically obscured by noise, precisely where entanglement is most robust.

To address this limitation, we construct a Hermitian generator for the QFI using only the Stokes contribution. Specifically, we consider an operator closely related to $\mathcal{T}_{{\bm{q}}}(\epsilon_{\rm i},\epsilon_{\rm s})$:
\begin{equation}\label{eq:HermitianGenerator} 
    \mathcal{O}_{{\bm{q}}}(\epsilon_{\rm i},\epsilon_{\rm s}, \varphi_{\bm{q}}) = \frac{1}{\sqrt{2}}\left[ e^{i\varphi_{\bm{q}}} \mathcal{T}_{{\bm{q}}}(\epsilon_{\rm i},\epsilon_{\rm s}) + e^{-i\varphi_{\bm{q}}} \mathcal{T}_{{\bm{q}}}^\dagger(\epsilon_{\rm i},\epsilon_{\rm s}) \right] ,
\end{equation}
which is Hermitian by construction. The QFI associated with this operator for a pure state takes the form
\begin{align} \label{eq:QFIgeneral} 
    F_{\mkern-2mu Q}\mkern-1mu[\mkern-1mu\mathcal{O}_{\bm{q}}(\epsilon_{\rm i}, \mkern-2mu \epsilon_{\rm s}, \mkern-2mu \varphi_{\bm{q}})\mkern-2mu] 
    \mkern-4mu = \mkern-4mu 2\langle \mathcal{T}^\dagger_{\bm{q}} \mathcal{T}_{\bm{q}} \rangle_c \mkern-4mu + \mkern-4mu 2\langle \mathcal{T}_{\bm{q}} \mathcal{T}_{\bm{q}}^\dagger \rangle_c \mkern-4mu + \mkern-4mu 4\mathrm{Re} \mkern-4mu \left[e^{2i\varphi_{\bm{q}}} \mkern-2mu \langle \mathcal{T}_{\bm{q}}^2 \rangle \mkern-3mu \right]\mkern-2mu.
\end{align} 
The first term corresponds exactly to the RIXS spectral weight defined in Eq.~\eqref{eq:RIXSIntegral}. The second term, $\langle \mathcal{T}_{\bm{q}} \mathcal{T}_{\bm{q}}^\dagger \rangle_c$, is not captured by Eq.~\eqref{eq:RIXSIntegral}, but becomes experimentally accessible by reversing momentum and exchanging the incident and scattered photon polarizations. This yields a scattering operator,
\begin{equation}
    \mathcal{T}_{-{\bm{q}}}(\epsilon_{\rm s},\epsilon_{\rm i}) = \sum_{j}e^{-i{\bm{q}} \cdot \bm{r}_j} \mathcal{D}_j^{(\epsilon_{\rm{i}})\dagger}\mathcal{D}_j^{(\epsilon_{\rm{s}})}
    = \mathcal{T}_{{\bm{q}}}^\dagger(\epsilon_{\rm i},\epsilon_{\rm s}) ,
\end{equation}
whose full integral provides the second term in Eq.~\eqref{eq:QFIgeneral}.

\begin{figure}[!t]
    \centering
   \includegraphics[width=8.5cm]{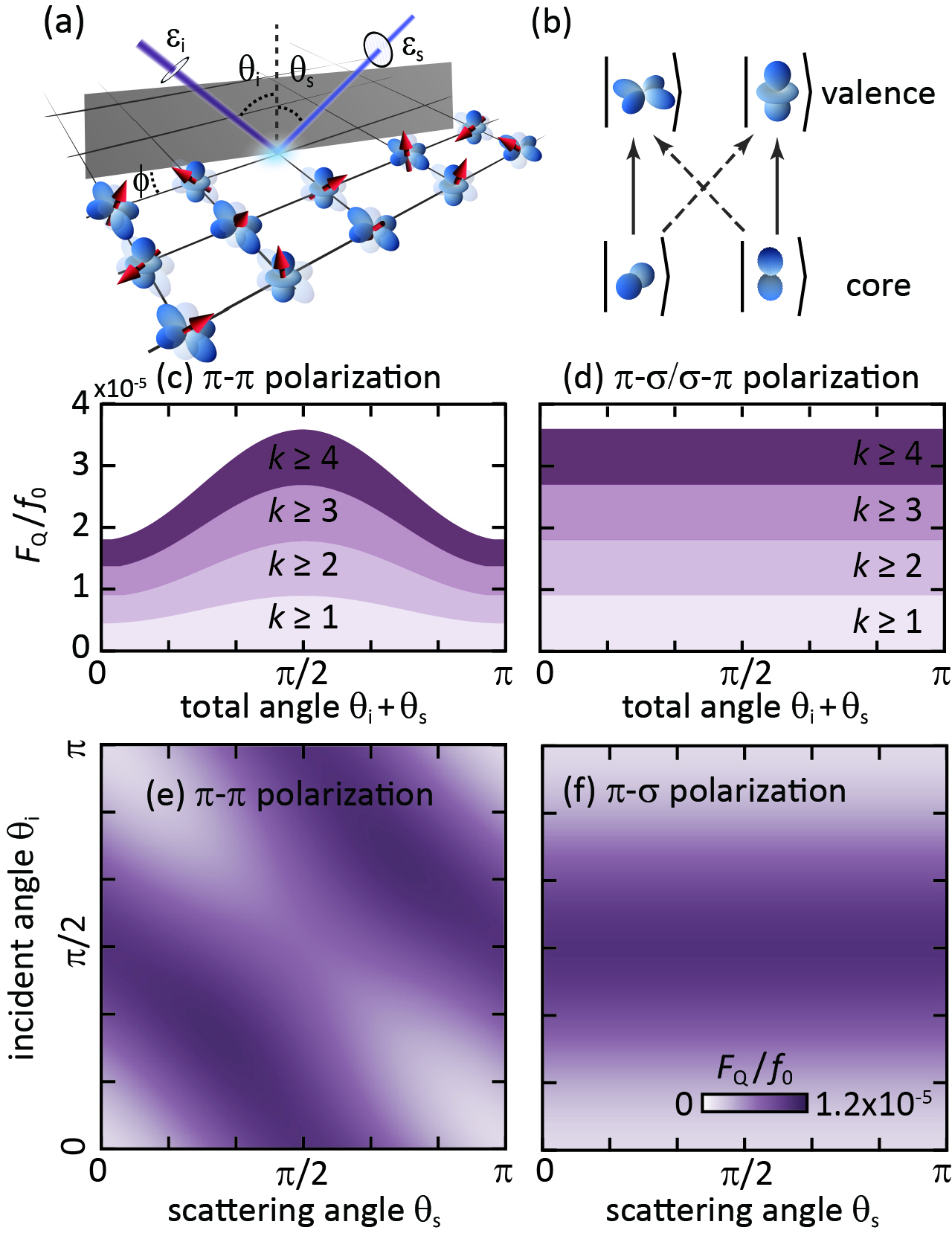}\vspace{-3mm}
    \caption{(a) RIXS experimental geometry. Incident and scattered photon wavevectors define the gray scattering plane. (b) X-ray-induced core-to-valence transitions. Solid arrows represent dominant dipole-allowed channels, while dashed arrows denote transitions enabled by crystal-field anisotropy. (c),(d) Dimensionless QFI bounds at fixed azimuthal angle $\phi = 0$, computed using spherically symmetric atomic orbitals for (c) $\pi$–$\pi$ and (d) $\pi$–$\sigma$ or $\sigma$–$\pi$ polarizations. (e),(f) Angular dependence of $k=1$ QFI bounds computed using molecular orbitals based on X2C-CASSCF for (e) $\pi$–$\pi$ and (f) $\pi$–$\sigma$ polarizations. }
    \label{fig:fixedPolarization}
\end{figure}

The third term, involving $\langle \mathcal{T}_{\bm{q}}^2 \rangle$, generally vanishes except at specific high-symmetry points such as ${\bm{q}} = (0,0)$ or $(\pi,\pi)$. To eliminate this term entirely, we exploit the phase degree of freedom in Eq.~\eqref{eq:HermitianGenerator}. Specifically, we select a Hermitian operator $\bar{\mathcal{O}}_{\bm{q}}(\epsilon_{\rm i},\epsilon_{\rm s})={\mathcal{O}}_{\bm{q}}\left[\epsilon_{\rm i},\epsilon_{\rm s}, \bar{\varphi}_{\bm{q}}(\epsilon_{\rm i},\epsilon_{\rm s})\right]$ such that
\begin{equation}
\bar{\varphi}_{\bm{q}}(\epsilon_{\rm i},\epsilon_{\rm s}) = \frac\pi 4 - \frac12\mathrm{Arg}\left( \langle \mathcal{T}_{\bm{q}}(\epsilon_{\rm i},\epsilon_{\rm s})^2 \rangle\right) .
\end{equation}
This choice nullifies the third term in Eq.~\eqref{eq:QFIgeneral}. With this optimized phase, the resulting QFI for $\bar{\mathcal{O}}_{\bm{q}}$ is then expressible purely through experimentally accessible RIXS spectra,
\begin{align}\label{eq:finalQFIIntegral}
    F_Q[\bar{\mathcal{O}}_{\bm{q}}(\epsilon_{\rm i},\epsilon_{\rm s})] \mkern-3mu= \mkern-3mu 2\Gamma^2 \mkern-6mu \int_{\mkern-1mu 0}^{\mkern-1mu \infty} \mkern-4mu \biggl [I({\bm{q}},\mkern-2mu\epsilon_{\rm i},\mkern-2mu\epsilon_{\rm s},\mkern-2mu\omega) \mkern-4mu + \mkern-4mu I({-\bm{q}},\mkern-2mu \epsilon_{\rm s},\mkern-2mu \epsilon_{\rm i},\mkern-2mu \omega)\biggl]\mkern-2mu \mathrm{d}\omega.
\end{align}
For a $k$-producible state $\ket{\Psi_{k-\rm{prod}}} $, it is bounded by\,\cite{hyllus2012fisher},
\begin{align}\label{eq:QFIbound}
    F_Q[ \bar{\mathcal{O}}_{\bm{q}}(\epsilon_{\rm i},\epsilon_{\rm s})]\leqslant k \sum_{j=1}^N \left[\lambda_{j}^{\rm (max)} - \lambda_{j}^{\rm (min)}\right]^2  ,
\end{align}
where $\lambda_{j}^{\rm (max)}$ and $\lambda_{j}^{\rm (min)}$ denote the largest and smallest eigenvalues of the Hermitian matrix $\bar{T}_{\alpha\beta}^{(j,{\bm{q}} )}(\epsilon_{\rm i},\epsilon_{\rm s})=e^{i ({\bm{q}} \cdot \bm{r}_j + \bar\varphi_{\bm{q}})} T_{\alpha\beta}(\epsilon_{\rm i},\epsilon_{\rm s}) + e^{-i ({\bm{q}} \cdot \bm{r}_j + \bar\varphi_{\bm{q}})} T_{\alpha\beta}(\epsilon_{\rm s},\epsilon_{\rm i})$, respectively. Violation of this inequality indicates at least $(k+1)$-partite spin-orbital entanglement. Unlike recent proposals relying on anti-Stokes contributions with Boltzmann weights\,\cite{ren2024witnessing}, our approach instead utilizes only the Stokes response ($\omega>0$), offering significantly enhanced robustness for low-temperature entanglement detection.

In contrast to spin systems, which often feature unified probe operators, the Hermitian operators and their associated QFI bounds in spin-orbital systems are intrinsically material dependent. In particular, the evaluation of the eigenvalue spread $\Delta \lambda_j = \lambda_{j}^{\rm (max)} - \lambda_{j}^{\rm (min)}$ requires explicit knowledge of the matrix $T_{\alpha\beta}(\epsilon_{\rm i}, \epsilon_{\rm s})$. This matrix originates from the dipole transition elements $M_{\alpha\beta}(\epsilon)$, which are computed via \textit{ab initio} single-electron integrals involving localized Wannier orbitals:
\begin{equation}
    M_{\alpha\beta}(\epsilon_{\rm i/s}) = e\sqrt{\frac{\hbar\omega_{\rm i/s}}{{2\varepsilon_0 V}}}\int d^{3}\bm{r}\, \psi_{\alpha}^{*}(\bm{r})\, (\bm\epsilon_{\rm i/s} \cdot \bm{r})\, \psi_{\beta}(\bm{r}) .
\end{equation}
Here, $\psi_\alpha(\bm{r})$ represents the localized Wannier wave functions, $\omega_{\rm i/s}$ is the photon frequency, $\varepsilon_0$ is the vacuum permittivity, and $V$ is the quantization volume. Although these dipole elements are frequently represented in the spin-orbit-coupled $j$ basis\,\cite{ament2009theoretical,haverkort2010theory}, the resulting eigenvalue spectrum remains invariant under basis transformations.

To exemplify the practical evaluation of QFI bounds, we consider cuprates, a canonical family of strongly correlated systems where low-energy excitations are dominated by the Cu $3d$ orbitals. Extensive studies have revealed active orbital excitations and their interplay with spins in cuprates\,\cite{schlappa2012spin,matt2018direct}, a topic of renewed interest due to emerging parallels in superconducting nickelates\,\cite{yang2024orbital,liu2024electronic,chen2024electronic}. The QFI bound in Eq.~\eqref{eq:QFIbound} depends on both the scattering momentum $\bm{q}$ and the polarization configurations. This structure naturally enables RIXS measurements performed across different momentum-polarization settings to serve as a sequence of independent entanglement probes, analogous to performing multiple Bell-type tests\,\cite{aspect1982experimental,brunner2014bell}. We adopt the scattering geometry depicted in Fig.~\ref{fig:fixedPolarization}(a), where the polarization vector $\epsilon $ is parameterized by the polar angle $\theta$ and the azimuthal angle $\phi$. In this fixed geometry, the momentum transfer $\mathbf{q}=\mathbf{k}_{\mathrm{in}}-\mathbf{k}_{\mathrm{out}}$ is uniquely determined by the incident and scattering angles $(\theta_i,\theta_s)$, with $|\mathbf{q}|=2k\sin[(\theta_i+\theta_s)/2]$ for $|\mathbf{k}_{\mathrm{in}}|\simeq|\mathbf{k}_{\mathrm{out}}|=k$. For each setting, we compute the dipole matrix elements $M^{(\epsilon_{\rm i})}_{\alpha\beta}$ corresponding to Cu $L$-edge transitions involving $2p$ core orbitals and use these to determine the eigenvalue spread $\Delta \lambda_j$. 

We first employ a spherically symmetric atomic orbital basis to estimate $M^{(\epsilon_{\rm i})}_{\alpha\beta}$. In this approximation, the wave functions $\psi_{\alpha}(\bm{r})$ for both core and valence orbitals are spherical harmonics that obey selection rules [see solid lines in Fig.~\ref{fig:fixedPolarization}(b)]. The matrix elements can then be evaluated analytically using the Wigner-Eckart theorem, which decomposes each element into a Clebsch-Gordan coefficient and a radial integral [see SM for derivations\,\cite{SI_note}]. 
For convenience, we express the QFI in a dimensionless form $F_Q / f_0$, with $f_0 = e^2\hbar^2 \omega_{\rm i}\omega_{\rm s}/4 \varepsilon_0^2 V^2$\,\cite{normalizationNote}. As shown in Fig.~\ref{fig:fixedPolarization}(c), when both polarizations lie in the scattering plane (the $\pi$-$\pi$ configuration), the QFI bounds vary strongly with the total angle $\theta_{\rm i} + \theta_{\rm s}$ between the beams, reaching a maximum when the incident and scattering beams are perpendicular. By contrast, rotating either polarization into the perpendicular direction (the $\pi$-$\sigma$ or $\sigma$-$\pi$ configuration) enforces a fixed relative polarization angle to $\pi/2$ regardless of the beam geometry. Owing to the spherical symmetry of the bases, this yields constant QFI bounds across all angles [see  Fig.~\ref{fig:fixedPolarization}(d)]. Notably, these values coincide with the maxima achieved in the $\pi$-$\pi$ configuration at orthogonal polarization. For a fixed geometry and polarization configuration, the QFI bounds for $k$-producible states are obtained by linear rescaling the $k=1$ bound according to Eq.~\eqref{eq:QFIbound}.

Estimates based on atomic orbitals assume idealized spherical symmetry, which breaks down in realistic crystalline environments. To better capture the electronic structure of cuprates, we perform high-precision quantum chemistry simulations of the CuO$_2$ plane using the exact two-component complete active space self-consistent field (X2C-CASSCF) method\,\cite{jenkins.2019.10.1021/acs.jctc.9b00011a}. This \textit{ab initio} simulation variationally incorporates spin-orbit and orbital anisotropy in a many-body context [see SM for details\,\cite{SI_note}]. 
Using this method, we recalculate the dipole transition operators $M^{(\epsilon_{\rm i})}_{\alpha\beta}$. As shown in Fig.~\ref{fig:fixedPolarization}(e), the resulting QFI bounds for the $\pi$-$\pi$ polarization channel now exhibit dependence on both $\theta_\mathrm{i}$ and $\theta_\mathrm{s}$, unlike the atomic bases, where they depend only on the sum of the two angles [see Fig.~\ref{fig:fixedPolarization}(c)]. Likewise, for the $\pi$–$\sigma$ (or $\sigma$–$\pi$) polarization channels, the QFI bounds are no longer uniform but instead vary with $\theta_\mathrm{i}$ (or $\theta_\mathrm{s}$). However, they remain insensitive to the angle associated with the $\sigma$-polarized beam since its polarization is independent from the angle [see Fig.~\ref{fig:fixedPolarization}(f)].

Each RIXS measurement [together with its conjugate defined in Eq.~\eqref{eq:finalQFIIntegral}], specified by a given polarization configuration and momentum, provides a lower bound on entanglement depth, and combining multiple measurements progressively tightens this bound. In practice, however, many beamlines cannot fully resolve photon polarizations, especially for the scattered photons entering spectrometers. Under such circumstances, the measured spectrum becomes an incoherent mixture of polarization channels, expressed as
\begin{equation}\label{eq:mixedPolIntensity}
I^{\mathrm{(\mpol)}}(\bm q, \omega) = \mkern-16mu\sum_{\epsilon_{\rm i},\epsilon_{\rm s}\in\{\pi, \sigma\}} \mkern-16mu w(\epsilon_{\rm i},\epsilon_{\rm s})\,  I({\bm{q}},\epsilon_{\rm i},\epsilon_{\rm s},\omega) .
\end{equation}
Here, $w(\epsilon_{\rm i},\epsilon_{\rm s})$ denotes the weight of each polarization configuration. Because these weights are \textit{a priori} unknown and generally do not satisfy reversal symmetry [i.e., $w(\epsilon_{\rm i},\epsilon_{\rm s}) \neq w(\epsilon_{\rm s},\epsilon_{\rm i})$], the conjugate-pair construction required for Eq.~\eqref{eq:finalQFIIntegral} is no longer applicable.

\begin{figure}[!t]
    \centering
    \includegraphics[width=\linewidth]{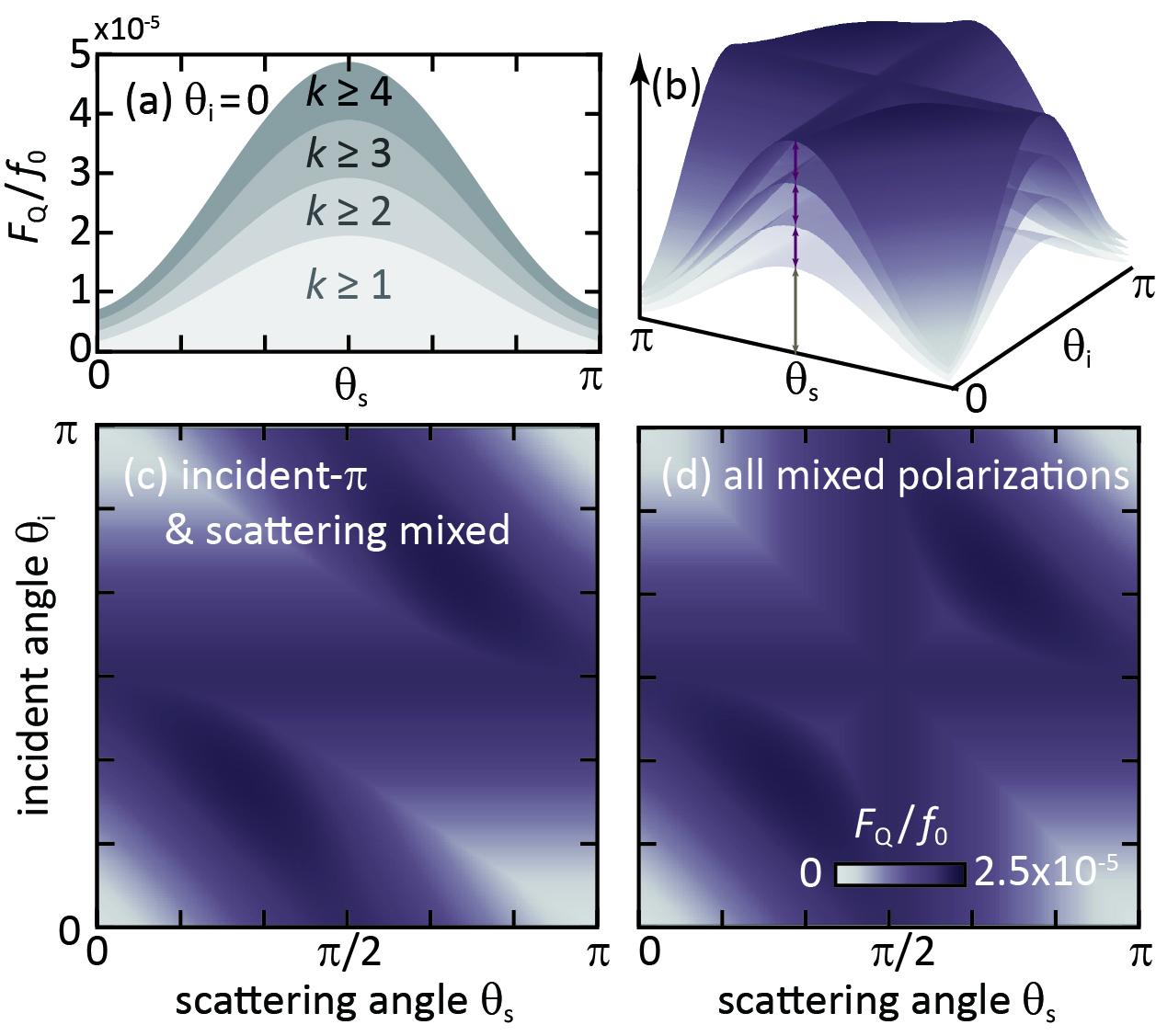}\vspace{-3mm}
    \caption{(a),(b) QFI bounds for $k$-producible states derived from mixed-polarization RIXS spectra for (a) the $\theta_{\mathrm i}=0$ cut and (b) the full scattering geometry. (c),(d) Angular dependence of the $k=1$ bound for (c) incident $\pi$-polarized and polarization-unresolved scattered beams and (d) fully unresolved polarization in both beams. Electronic structure inputs are obtained via X2C-CASSCF calculations.}
    \label{fig:mixedPolarization}
    \vspace{-1pt}
\end{figure}

Even without a directly accessible Hermitian operator for constructing QFI, the integral of the mixed-polarization RIXS spectra still admits an upper bound by a crude estimate [see detailed derivations in SM\,\cite{SI_note}]:
\begin{align}\label{eq:mixedPolQFI}
    4\Gamma^2\mkern-6mu\int_0^\infty \mkern-8mu  &I^{\mathrm{(mp)}}(\mathbf q,\omega)\,{\rm d}\omega
    \leqslant \max_{\epsilon_{\rm i},\epsilon_{\rm s}}\{F_Q[  \bar{\mathcal{O}}_{\bm{q}}(\epsilon_{\rm i},\epsilon_{\rm s})]\}\nonumber\\
     &\quad\quad+ 2N\,\max_{\epsilon_{\rm i},\epsilon_{\rm s}}\{\lambda_{\max}\big([T^\dagger(\epsilon_{\rm i},\epsilon_{\rm s}), T(\epsilon_{\rm i},\epsilon_{\rm s})]\big)\},
\end{align}
where $\lambda_{\max}(A)$ denotes the largest eigenvalue of matrix $A$. The first term on the right side of Eq.~\eqref{eq:mixedPolQFI} represents the convex envelope of the polarization-resolved QFI bounds from Eq.~\eqref{eq:QFIbound} and thus scales as $\order{kN}$ for a $k$-producible state, providing spin-orbital entanglement witness. The second term in Eq.~\eqref{eq:mixedPolQFI} produces an offset linear in system size $N$ but independent of the entanglement depth $k$. Although this contribution cannot be experimentally removed in mixed-polarization RIXS, its magnitude can be computed accurately for specific materials from the single-site commutator $[T^\dagger,T]$ using quantum chemistry.

Figure \ref{fig:fixedPolarization} presents the bounds for $k$-producible states obtained by mixed-polarization RIXS. The angular dependence of the resulting bounds is shown in Figs.~\ref{fig:mixedPolarization}(a) and \ref{fig:mixedPolarization}(b), which display a representative $\theta_{\rm i} = 0$ cut and the full scattering geometry, respectively. In comparison with the polarization-resolved QFI bounds in Fig.~\ref{fig:fixedPolarization}, the mixed-polarization bounds are substantially looser at all entanglement depths due to the loss of polarization specificity. Although more conservative manifesting as an offset for the $k=1$ upper bounds, the hierarchy of bounds retain the relative separation between $k$-producible states, indicating that the mixed-polarization protocol remains sensitive to multipartite entanglement. 

To explore this further, we examine two practical scenarios: one in which the incident beam is fixed to $\pi$ polarization and one in which both incident and scattered polarizations are unresolved. The corresponding $k=1$ bounds, shown in Figs.~\ref{fig:mixedPolarization}(c) and \ref{fig:mixedPolarization}(d), reveal new regions of enhanced intensity, e.g., near $\theta_{\rm i} = \theta_{\rm s} = \pi/2$. These regions do not necessarily correspond to strong RIXS intensity, but instead arise from the geometry’s insensitivity to polarization mixing, illustrating how unresolved polarizations obscure entanglement information. Therefore, for best results under polarization-unresolved conditions, it is advantageous to select scattering geometries for which the mixed-polarization bounds in Figs.~\ref{fig:mixedPolarization}(c) and ~\ref{fig:mixedPolarization}(d) most closely approximate the polarization-resolved bounds in Figs.~\ref{fig:fixedPolarization}(e) and \ref{fig:fixedPolarization}(f). Such geometries offer the tightest bound conditions for quantifying entanglement depth under mixed-polarization RIXS measurements.

In summary, we present a RIXS-based strategy to detect spin-orbital entanglement in quantum materials by constructing Hermitian generators from non-Hermitian scattering operators using paired conjugate geometries. This framework yields scattering geometry- and polarization-resolved bounds for $k$-producible states, determined by the local eigenvalue spread of the scattering matrix elements. We further extend this approach to experimentally relevant scenarios where polarization cannot be resolved. In this case, we obtained a more conservative bound than fully polarization-resolved bounds, that relies on a single polarization-unresolved RIXS spectrum. Together, these results establish a practical hierarchy of QFI-based entanglement witnesses applicable to a broad range of RIXS measurements. Our results significantly broaden the scope of RIXS-based entanglement witnessing beyond magnetically ordered systems.

\textit{Acknowledgments} - We acknowledge insightful discussions from Robert Konik, Matteo Mitrano, Yingying Peng, and Sophia TenHuisen. This work is primarily supported by the U.S. Department of Energy, Office of Science, Basic Energy Sciences, under Early Career Award No.~DE-SC0024524. Z.Z. and F.A.E acknowledge the support of the U.S. Department of Energy, under Award No.~DE-SC0024532. S. D. is also partially supported by the Scialog Grant No. SA-QMI-2025-083a from Research Corporation for Science Advancement and Kevin Wells. The simulation used resources of the National Energy Research Scientific Computing Center, a U.S. Department of Energy Office of Science User Facility located at Lawrence Berkeley National Laboratory, operated under Contract No.~DE-AC02-05CH11231 using NERSC Award No. BES-ERCAP0031226.

\textit{Data Availability} - The data that support the findings of this article are openly available. \,\cite{figshare}.

\bibliography{references}
\end{document}